\documentclass[prd,preprint,showpacs,aps,epsfig]{revtex4}
\usepackage{latexsym}
\usepackage{color}
\usepackage{graphicx}      
\usepackage{subfig}
\usepackage{amsmath,amssymb}
\usepackage{makeidx}
\usepackage{hyperref}       
\usepackage{subeqnarray}
\usepackage[utf8]{inputenc} 
\usepackage{indentfirst}
\usepackage{booktabs}
\usepackage{multirow}
\usepackage{array}

\graphicspath{{Figuras/}}

\makeindex

\begin{document}
	
\title{Gravitational collapse for a radiating anisotropic fluid}
	
\author{L. S. M. Veneroni}
\email{leone\_melo@yahoo.com.br}
\author{M. F. A. da Silva}
\email{mfasnic@gmail.com}
	
\affiliation{
		Departamento de F\' {\i}sica Te\' orica, Universidade do Estado do Rio de Janeiro, Rua S\~ ao Francisco Xavier $524$, Maracan\~ a, CEP 20550--013, Rio de Janeiro, RJ, Brazil}
	
\date{\today}
	
\pacs{xxxx}
	
\begin{abstract}
Interested in the collapse of a radiating star, we study the temporal evolution of a fluid with heat flux and bulk viscosity, including anisotropic pressure.
As a starting point, we adopt an initial configuration that satisfies the regularities conditions as well as the energy conditions to a certain range of the mass-radius ratio for the star, defining acceptable models. For this set of models, we verify that the energy conditions remain satisfied until the black hole formation. Astrophysical relevant quantities, such as the luminosity perceived by an observer at infinity, the time of event horizon formation and the loss of mass during the collapse are presented.  

\end{abstract}


\maketitle

\section{Introduction}

Understanding the processes that lead to gravitational collapse, forming black holes or naked singularity (in the context of the cosmic sensorship), or even the gravitational stability allowing the description of stars in the point of view of general relativity, are astrophysical problems that are still far from properly approach.

Although many analytical solutions can be found, only a small number of them represent physically acceptable stellar models, even though they are far from providing a realistic description. From the pioneer model of a  star proposed by Oppenheimer and Snyder \cite{OS1939} for neutral matter, and even the Schwarzschild interior solution considering a perfect fluid with constant energy density \cite{Schw1916b}, many researchers have strived to search for more realistic models. A first step of great importance in this direction was given by Vaidya in 1951 \cite{V1951}, when he obtained the first exact solution to space time filled by null radiation, with spherical symmetry, and thus representing the outer gravitational field of a star radiant. From there, solutions considering the presence of heat flux in the stellar interior began to be investigated. However, only in 1985, with the work of Santos \cite{S1985} proposing the conditions of connection between a source solution with radial heat flow and an exterior represented by the Vaidya solution, these studies could bring more significant results. 

Several authors attempted to construct more realistic models and to understand how the stellar interior behaves, such as Oliveira et al. \cite{OKS1988}, Kolassis et al. \cite{KST1988}, and Bonnor et al. \cite{BON1989} in early years. 

Another contribution to this subject was done by Chan, who initially considered only anisotropic stresses \cite{C1993}, after that, he included the contributions of the shear \cite{C1997}\cite{C1998}\cite{C2000} and the  expasion scalar \cite{C2001} in the stress-energy tensor. Posteriorly, Nogueira and Chan \cite{NC2004}, and Pinheiro and Chan \cite{PC2010}  continued these studies.    

Herrera and Santos introduced the concept of euclidean stars \cite{HS2010}, simplifying the metric in order to introduce non null shear in an exact solution. This kind of source has been explored in G. Govender et al. \cite{GGG2010}, Govinder and M. Govender \cite{GG2012}, and Abede et al. \cite{AMG2014}. Recently, M. Govender et al. also studied the effect of shear in a dissipative gravitational collapse \cite{GRM2014}. 

In the gravitational collapse context, causal thermodynamics was introduced by M. Govender et al. \cite{GMM1998}. Other works also included it, such as Naidu et al. \cite{NGG2006}, M. Govender et al. \cite{GRM2014}, and Naidu and M. Govender \cite{NG2016}.  

The first exact solution for an radiating star was obtained by Naidu et al. \cite{NGG2006}. This work was genelarized by Rajah and Maharaj \cite{RM2008} when they treated juction condition as a Riccati equation.

In this work, we start from a static initial solution for an anisotropic fluid in the pressures, obtained first by Hernández and Núñez \cite{HN2004} under the hypothesis of a non-local state equation \cite{HNP1999} and from the choice of the energy density profile of Gokhroo and Mehra \cite{GM1994}. It is important to note that this solution  satisfies all conditions of regularity, hydrodynamic stability, as well as energy, and the latter limits the possible stellar models as a function of their mass-radius ratio. In order to study the temporal evolution of a plausible initial configuration such as this, we introduce a temporal dependence on the metric and take the static solution as the initial limit of the collapse process.  
  
Our paper is organized as follows. In section 2, we present the Einstein's field equations. Once that is done, we introduce the junction conditions in section 3, with Vaidya's metric describing the exterior space-time, just as some relevant physical quantities. Following that, in section 4, we propose a solution for the field equations. In section 5, the initial static solution is presented in details.considering a static density profile, we can find the anisotropic pressures. Then, in section 6, we investigate if a black hole is created, when it occurs, as well as the values of the effective adiabatic index along the star, during the collapse for different values of the bulk viscosity coefficient. We also obtained the luminosity for a distant observer, and how much mass is loss due to radiation process. In section 7, we verify that there are configurations which satisfy all the  energy conditions inside the source during the collapse process. Finally, we make the concluding remarks in section 8.

\section{Einstein field equations}

Our purpose in this work is to study the behavior of a dissipative fluid distribution, initially static and sufficiently compacted to justify the use of general relativity. In other words, we are looking for the modeling of an astrophysical object, such as a neutron star for example, that is allowed to collapse and follow the evolution of some physical parameters relevant to the model. The most general spherically symmetric matter distribution for inner spacetime, corresponding to the source, can be written as

\begin{eqnarray}
\label{met3}
ds^2_{-} &=& g^{-}_{\alpha\beta}d\chi^{\alpha}_{-}d\chi^{\beta}_{-} \nonumber\\
&=& -A^{2}(r,t)dt^{2}+B^{2}(r,t)dr^2+C^{2}(r,t)\left(d\theta^2+\sin^{2}(\theta)d\phi^2\right) \,.
\end{eqnarray}

The energy momentum tensor representing the dissipative fluid, including terms, is given by

\begin{eqnarray}
\label{tens1}
T_{\alpha\beta}^{-} = (\rho + P_{\perp})u_{\alpha}u_{\beta}+P_{\perp}g_{\alpha\beta}+(P_{r}-P_{\perp})X_{\alpha}X_{\beta} \nonumber\\
+q_{\alpha}u_{\beta}+q_{\beta}u_{\alpha}-2\eta\sigma_{\alpha\beta}-\zeta\Theta(g_{\alpha\beta}+u_{\alpha}u_{\beta}),
\end{eqnarray}
where $\eta>0$ and $\zeta>0$ are the coefficients of shear viscosity and bulk viscosity, respectively. The quantity $\rho$ is the energy density, while $P_r$ is the radial pressure, $P_\perp$ is the tangential one. Besides, $\sigma_{\alpha\beta}$ is the shear tensor, $\Theta$ is the expansion scalar, $q^\alpha$ is the heat flow, $u^\alpha$ is the four-velocity, and $X^\alpha$ is an unit vector. The last vectors should satisfy the relations $u^{\alpha}u_{\alpha}=-1$, $u^{\alpha}q_{\alpha}=0$, $X_{\alpha}X^{\alpha}=1$ and $X_{\alpha}u^{\alpha}=0$, being defined as $u^\alpha=\delta^\alpha_0/A$, in an inertial frame of reference comoving with the fluid, $q^\alpha=q\delta^\alpha_1$ and $X^\alpha=\delta^\alpha_1/B$.

Thereby, combining the metric (\ref{met3}) with the energy momentum tensor (\ref{tens1}), the Einsteins's field equations provide us

\begin{eqnarray}
\label{eins1}
-\left(\frac{A}{B}\right)^{2}\left[2\frac{C''}{C}+\left(\frac{C'}{C}\right)^{2}-2\frac{C'}{C}\frac{B'}{B} \right]+ \left(\frac{A}{C}\right)^{2}+ \frac{\dot{C}}{C}\left(\frac{\dot{C}}{C}+2\frac{\dot{B}}{B}\right) \nonumber\\
=\kappa A^{2}\rho \, ,
\end{eqnarray} 

\begin{eqnarray}
\label{eins2}
\frac{C'}{C}\left(\frac{C'}{C}+2\frac{A'}{A}\right) - \left(\frac{B}{C}\right)^2  - \left(\frac{B}{A}\right)^{2}\left[2\frac{\ddot{C}}{C}+\left(\frac{\dot{C}}{C}\right)^{2}-2\frac{\dot{A}}{A}\frac{\dot{C}}{C} \right] \nonumber\\
= \kappa B^2(P_r+4\eta\sigma-\zeta\Theta) \,,
\end{eqnarray} 

\begin{eqnarray}
\label{eins3}
\left(\frac{C}{B}\right)^{2}\left[\frac{A''}{A}+\frac{C''}{C}- \frac{A'}{A}\frac{B'}{B}+\frac{A'}{A}\frac{C'}{C} -\frac{B'}{B}\frac{C'}{C}\right] \nonumber\\
+\left(\frac{C}{A}\right)^{2}\left[-\frac{\ddot{B}}{B}-\frac{\ddot{C}}{C}+\frac{\dot{A}}{A}\frac{\dot{B}}{B}+\frac{\dot{A}}{A}\frac{\dot{C}}{C}-\frac{\dot{B}}{B}\frac{\dot{C}}{C}\right] \nonumber\\ 
= \kappa C^2(P_\perp-2\eta\sigma-\zeta\Theta) \, , 
\end{eqnarray} 

\begin{equation}
\label{eins5}
2\frac{A'}{A}\frac{\dot{C}}{C}+2\frac{\dot{B}}{B}\frac{C'}{C}-2\frac{\dot{C'}}{C}=-\kappa AB^2q \,,
\end{equation}\\*
with $\kappa =8\pi$ in the geometric coordinate system $(c=G=1)$, which means that mass, length and time have the same dimensions. Here we adopted all of them  in seconds. The line represents  $\partial/\partial r$ and the dot means $\partial/\partial t$.  

\section{Junction conditions}

We separate the manifolds into interior and exterior by a timelike three-space spherical hipersurface $\Sigma$. The interior spacetime, representing a star, is fullfiled by the dissipative and anisotropic fluid described by the energy-momentum tensor, given by  (\ref{tens1}). Since we are considering a heat flow in the interior of the star, in order to become the process more realistic, it is reasonable to suppose that the spherical distribution of matter emits null radiation while collapses. Then, the exterior spacetime is not vacuum but fullfiled by null radiation. The more general metric which represents the solution of the Einstein's equations  describing a spherically symmetrical distribution of this type of radiation is known as Vaydia's metric and it is given by

\begin{eqnarray}
\label{met4}
ds^2_{+} &=& g^{+}_{\alpha\beta}d\chi^{\alpha}_{+}d\chi^{\beta}_{+} \nonumber\\
&=& -\left(1-\frac{2m(v)}{\mathbf{r}}\right)dv^{2}-2dvd\mathbf{r}+\mathbf{r^2}(d\theta^2+\sin^{2}(\theta) d\phi^2) \,,
\end{eqnarray}
where $\chi^{\alpha}_{+}=(\chi^{0}_{+},\chi^{1}_{+},\chi^{2}_{+},\chi^{3}_{+})=(v,\mathbf{r},\theta,\phi)$.\\

It is important to note that the null radiation is described by the energy-momentum tensor of a null "dust" fluid, that is, a perfect fluid with zero pressure and energy density given by

\begin{equation}\label{nullrad}
\mu(\mathbf{r},v) = -\frac{1}{\mathbf{r}^2}\dfrac{dm(v)}{dv} \,,
\end{equation}\\
which implies that $dm(v)/dv$ must always be negative, that is, the star will always lose energy, never absorbing it.

To ensure a smooth transition in the hypersurface between the two spacetimes, some conditions of continuity, known as junction conditions, have been established, imposing that the first and second fundamental forms are continuous (the metric and the extrinsic curvature, respectively). Following Nogueira and Chan \cite{NC2004}, the continuity of the first fundamental form provides us

\begin{equation}
\label{igual1}
\left(\frac{dt}{d\tau}\right)_\Sigma = \frac{1}{A(r_{\Sigma},t)} \,,
\end{equation}

\begin{eqnarray}
\label{igual5}
\left(\frac{dv}{d\tau}\right)_\Sigma & = & \left[\frac{1}{\sqrt{ 1-\frac{2m(v)}{\mathbf{r}}+\frac{2d{\mathbf{r}}}{d{v}}}}\right]_{\Sigma}\\ \nonumber
& = & \left[ \frac{-\frac{d{\mathbf{r}}}{d\tau}\pm\sqrt{\left( \frac{d{\mathbf{r}}}{d\tau}\right)^2 +1 -\frac{2m}{\mathbf{r}}}}{1-\frac{2m}{\mathbf{r}}}\right]_\Sigma \,,
\end{eqnarray}

\begin{equation}
\label{igual3}
C(r_{\Sigma},t) = \mathbf{r}_{\Sigma}(v) \,.
\end{equation}\

On the other hand, the continuity of the second fundamental form allows us to obtain the mass-energy function in the form

\begin{equation}
\label{massa3}
m = \left\lbrace \frac{C}{2}\left[1+\left(\frac{\dot{C}}{A}\right)^{2} - \left(\frac{C'}{B}\right)^{2}\right]\right\rbrace _\Sigma \,,   
\end{equation}\\*
where $m$ is the mass-energy contained within the distribution entrapped between the center and $\Sigma$. Note that this expression is equivalent to the definition given by Cahill and McVittie \cite{CM1970}.

In addition, the gravitational redshift can be put in the form 

\begin{equation}
\label{red2}
\left(\frac{dv}{d\tau}\right)^{-1}_\Sigma=\frac{1}{1+z_\Sigma} \,= \left(\frac{C'}{B}+\frac{\dot{C}}{A}\right)_{\Sigma} \,,
\end{equation}\\*
and it is simple to see that it diverges when

\begin{equation}
\label{red3}
\left(\frac{C'}{B}+\frac{\dot{C}}{A}\right)_{\Sigma} = 0 \,.
\end{equation}\

The continuity of the extrinsic curvature allows us to write the identity

\begin{equation}
\label{igual11}
\left(-\frac{C}{2AB}G^-_{01}\right)_\Sigma =
\left(\frac{C}{2B^2}G^-_{11}\right)_\Sigma \,.
\end{equation}\

If we combine (\ref{igual11}) with (\ref{eins2}) and (\ref{eins5}), we can find the following relation

\begin{equation}
\label{jun1}
\left(P_r+4\eta\sigma-\zeta\Theta\right)_\Sigma =
\left(qB\right)_\Sigma \,,
\end{equation}\\*
which is the generalization of the junction condition $\left(P_r\right)_\Sigma=0$ in the presence of dissipation and viscosity terms. 

Moreover, the total luminosity for an observer in rest (an adimensional quantity for the geometric units used here is given by

\begin{equation}
\label{lum1}
L = -\left(\frac{dm}{dv}\right)_\Sigma = -\left[\frac{dm}{dt}\frac{dt}{d\tau}\left(\frac{dv}{d\tau}\right)^{-1}  \right]_\Sigma \,. 
\end{equation}\

So, using (\ref{igual1}), (\ref{massa3}), (\ref{red2}) and (\ref{jun1}), we can rewrite (\ref{lum1}) as

\begin{equation}
\label{lum6}
L = \frac{\kappa}{2}\left[C^2\left(P_r+4\eta\sigma-\zeta\Theta\right)\left(\frac{\dot{C}}{A}+\frac{C'}{B}\right)^2\right]_\Sigma \,.
\end{equation}\

\section{The field equations solutions}

In order to study the evolution of an initial static configuration simulating a gravitational collapse, we consider the metric

\begin{equation}
\label{met12}
ds^2_{-}=-\frac{\xi^2}{h(r)}dt^2+\frac{f(t)}{h(r)}dr^2+f(t)r^2d\Omega^2\, ,
\end{equation}\\
where $\xi$ is an arbitrary constant, and the metric form was chosen so that it coincides with the metric proposed by Hernández e Núñez to a static configuration \cite{HN2004} when $f(t_0)= 1$, in such way that $t_0$ represents the initial time of the collapse. 

The Einstein's field equations (\ref{eins1})-(\ref{eins5}) for this metric reduce to

\begin{equation}
\label{eins6b}
\frac{\xi^2 \left( 1-h-rh'\right) }{r^{2}hf}+\frac{3}{4}\frac{\dot{f^2}}{f^2} = \frac{\xi^2 \rho}{h} \,,
\end{equation} 

\begin{equation}
\label{eins7b}
\frac{h-rh'-1}{r^{2}h}+\frac{1}{4\xi^2}\frac{\dot{f^2}}{f}-\frac{\ddot{f}}{\xi^2}= \frac{f}{h}(P_r+4\eta\sigma-\zeta\Theta) \,,
\end{equation} 

\begin{equation}
\label{eins8b}
r^2 \left[ \frac{h'^{2}-hh''}{2h}+\frac{h}{4\xi^2}\frac{\dot{f^2}}{f}- \frac{h}{\xi^2}\ddot{f} \right]= r^{2}\,f\,(P_\perp-2\eta\sigma-\zeta\Theta) \,,
\end{equation}

\begin{equation}
\label{eins10b}
 -\frac{h'}{2\,h}\frac{\dot{f}}{f}= -\frac{\xi f q}{\sqrt{h^3}} \,.
\end{equation}\ 

We can write the expansion scalar, the shear tensor and the shear scalar, respectively, as 

\begin{equation}
\label{Theta0}
\Theta = u^\alpha_{;\alpha} = \frac{1}{A}\left(\frac{\dot{B}}{B}+\frac{2\dot{C}}{C}\right)= \frac{3}{2\xi}\frac{\sqrt{h}\dot{f}}{f}  \, ,
\end{equation}

\begin{equation}
\label{sigma0}
\sigma_{\alpha\beta}=u_{(\alpha;\beta)}+\dot{u}_{(\alpha}u_{\beta)}-\frac{1}{3}\Theta (g_{\alpha\beta}+u_{\alpha}u_{\beta})\, ,
\end{equation}

\begin{equation}
\label{sigma4}
\sigma=-\frac{1}{\sqrt{6}}\sqrt{\sigma^{\alpha\beta}\sigma_{\alpha\beta}}=-\frac{1}{3A}\left(\frac{\dot{B}}{B}-\frac{\dot{C}}{C}\right) = 0\,,
\end{equation}
since $\dot B/B = \dot C/C$, and $\dot{u}_{\alpha}=u_{\alpha;\beta}u^\beta$.

So, the junction condition (\ref{jun1}) takes the following form

\begin{equation}
\label{jun1b}
\left( P_{r}-\zeta\Theta\right)_\Sigma = \left(qB\right)_\Sigma \,.
\end{equation}\

Considering the metric (\ref{met12}) in the mass-energy  function (\ref{massa3}), we have

\begin{equation}
\label{massa7}
m = \left\lbrace \frac{r\,\sqrt{f}}{2}\left[1+\frac{r^2h\dot{f}^2}{4\xi^2 f}-h\right] \right\rbrace _\Sigma \,.
\end{equation}\

From (\ref{eins6b})-(\ref{eins10b}), we can obtain the heat flux, the energy density, the radial and tangential pressure, respectively

\begin{equation}
\label{calor2}
8\pi q = \frac{\sqrt{h}\,h'}{2\xi}\frac{\dot{f}}{f^2} \,,
\end{equation}

\begin{equation}
\label{eins18}
8\pi\rho= \frac{1-rh'-h}{r^{2}f}+\frac{3h}{4\xi^2}\frac{\dot{f^2}}{f^2} \,,
\end{equation} 

\begin{equation}
\label{eins19}
8\pi (P_r-\zeta\Theta)= \frac{h-rh'-1}{r^{2}f}+\frac{h}{4\xi^2}\frac{\dot{f^2}}{f^{2}}-\frac{h}{\xi^2}\frac{\ddot{f}}{f}\,,
\end{equation} 

\begin{equation}
\label{eins20}
8\pi (P_\perp-\zeta\Theta)= \frac{h'^{2}-hh''}{2hf}+\frac{h}{4\xi^2} \frac{\dot{f^2}}{f^{2}}-\frac{h}{\xi^2}\frac{\ddot{f}}{f} \,.
\end{equation}\ 

The luminosity for an observer at rest at infinity can be obtained from (\ref{lum6}) and  (\ref{eins19}), that is,

\begin{equation}
\label{lum8}
L = \left\lbrace \frac{h}{8\xi^2 f}\left[h-rh'-1+\frac{r^2h\dot{f}^2}{4\xi^2 f}-\frac{r^2h\ddot{f}}{\xi^2} \right]\left[ r\dot{f}+2\xi\sqrt{f} \right]^2 \right\rbrace _\Sigma \,.
\end{equation}\

Then, using (\ref{met12}), (\ref{calor2}), (\ref{eins19}) into (\ref{jun1b}), we have got

\begin{equation}
\label{jun5}
\frac{h(r_\Sigma)-r_\Sigma h'(r_\Sigma)-1}{r^{2}_\Sigma h(r_\Sigma)}+\frac{1}{4\xi^2}\frac{\dot{f^2}}{f}-\frac{\ddot{f}}{\xi^2} = \frac{1}{2\xi}\frac{h'(r_\Sigma)}{h(r_\Sigma)}\frac{\dot{f}}{\sqrt{f}} \,.
\end{equation}\

At the initial time, the equation (\ref{jun5}) reproduces the static case from Hernández e Núñez \cite{HN2004}, that is

\begin{equation}
\label{jun6}
h'(r_\Sigma)=-\frac{1-h(r_\Sigma)}{r_\Sigma} \,.
\end{equation}\

So, substituting (\ref{jun6}) into (\ref{jun5}), we find 

\begin{equation}
\label{jun7}
\ddot{f}-\frac{1}{4}\frac{\dot{f^2}}{f}-\xi\left( \frac{1-h(r_\Sigma)}{2r_\Sigma\,h(r_\Sigma)}\right) \frac{\dot{f}}{\sqrt{f}} = 0 \,,
\end{equation}\\*
which can be rewrite as

\begin{equation}
\label{jun8}
\tilde{y}-\frac{y}{4f}-\xi\left(\frac{1-h(r_\Sigma)}{2r_\Sigma\, h(r_\Sigma)}\right)\frac{1}{\sqrt{f}} = 0 \,,
\end{equation}\\*
where $\tilde{y}=\dfrac{dy}{df}$.\\

Solving the equation (\ref{jun8}), we get

\begin{equation}
\label{fd1}
\dot{f}=2\xi\left( \frac{1-h(r_\Sigma)}{r_\Sigma\, h(r_\Sigma)}\right)f^{1/2}+af^{1/4} \,,
\end{equation}\\*
where $a$ is an arbitrary integration constant.\\

In order to determine $a$, we assume that $f \rightarrow 1$ and $\dot{f} \rightarrow 0$ when $t=t_0 \rightarrow -\infty$ (initial time). Thus,

\begin{equation}
\label{fd2}
a = -2\xi\left( \frac{1-h(r_\Sigma)}{r_\Sigma\, h(r_\Sigma)}\right) \,.
\end{equation}\

Using (\ref{fd2}), we rewrite (\ref{fd1}) in a simple form as

\begin{equation}
\label{fd3}
\dot{f}=-a\left(f^{1/2}-f^{1/4}\right)  \,.
\end{equation}\

Integrating the above equation, we obtain 

\begin{equation}
\label{t1}
t-t_*=-\frac{2}{a}\left( f^{1/2}+2f^{1/4}+2\ln|f^{1/4}-1|\right) \,,
\end{equation}\\*
where $t_*$ is an arbitrary constant.\\ 

The function $f$ has values between $1$ and $0$, so the equation (\ref{t1}) can be rewritten as

\begin{equation}
\label{t2}
t=-\frac{2}{a}\left[ f^{1/2}+2f^{1/4}+2\ln(1-f^{1/4})\right] \,,
\end{equation}\\*
where $t-t_* \rightarrow t$.\\

Since $-\infty < t \le 0$, we have

\begin{equation}
\label{t3}
0 < 1-f^{1/4} \le 1 \qquad \Rightarrow \qquad 0 \le f < 1 \,,
\end{equation}\\*
with $f \rightarrow 1$ when $t\rightarrow - \infty$.\\ 

Thus, the field equations (\ref{calor2}), (\ref{eins18}), (\ref{eins19}), and (\ref{eins20}), can be written in function of $f$, that is, 

\begin{equation}
\label{calor3}
8\pi q = b\,\sqrt{h}\,h'\left(\frac{f^{1/2}-f^{1/4}}{f^2}\right) \,,
\end{equation}

\begin{equation}
\label{eins21}
8\pi\rho = \frac{1-rh'-h}{r^{2}f}+3b^2h\left( \frac{f^{1/2}-f^{1/4}}{f}\right)^2 \,,
\end{equation} 

\begin{equation}
\label{eins22}
8\pi (P_r-\zeta\Theta)= \frac{h-rh'-1}{r^{2}f}+b^2h
\left(\frac{f^{-1/4}-1}{f}\right) \,,
\end{equation} 

\begin{equation}
\label{eins23}
8\pi (P_\perp-\zeta\Theta)=\frac{h'^{2}-hh''}{2hf} +b^2h\left(\frac{f^{-1/4}-1}{f}\right) \,,
\end{equation}
with $b = -\frac{a}{2\xi}$.\\ 

In the same way, using (\ref{fd3}), we can write (\ref{Theta0}) as

\begin{equation}
\label{Theta3}
\Theta = 3b\sqrt{h}\left(\frac{f^{1/2}-f^{1/4}}{f}\right) \,.
\end{equation}\

Substituting (\ref{fd3}) into (\ref{massa7}), we obtain

\begin{equation}
\label{massa8}
m=\frac{1}{2}\left\lbrace r\,\sqrt{f}\left[1+\frac{(1-h)^2}{h}\frac{\left( f^{1/2}-f^{1/4}\right)^2}{f}-h\right] \right\rbrace _\Sigma \,.
\end{equation}\

Next, using (\ref{jun6}) and (\ref{fd3}), we can write (\ref{lum8}) as

\begin{equation}
\label{lum9}
L = \frac{1}{2}\left\lbrace 
\left(1-h\right)^2\left(f^{-1/4}-1\right) \left[ 1 +\frac{\left(1-h\right)}{h}\left( 1-f^{1/4}\right) \right]^2 \right\rbrace _\Sigma \,.
\end{equation}\

\section{The initial static solution}

The initial configuration adopted in this work is one of that used by Hernández and Núñez \cite{HN2004}, using the Gokhroo and Mehra's density profile \cite{GM1994},

\begin{equation}
\label{rho1}
\rho (r)=\frac{9M_0}{r^3_\Sigma}\left(1-\frac{5r^2}{9r_{\Sigma}^2}\right) \,,
\end{equation}
being $M_0$ the initial mass, and $\rho$ the energy density given in $s^{-2}$, in geometric units.\\

We obtain the metric which describes the static solution from (\ref{met12}) if we impose that $f=1$ at the initial time. In this case, the exterior spacetime is described by the Schwarzschild metric. So, in order to match the inner and outer spacetimes, we have $h(r_\Sigma)=\xi$.
Thus, from the equations (\ref{eins21})-(\ref{Theta3}), we obtain

\begin{equation}
\label{h1}
h(\delta) = 1-\gamma\delta^2 \left(3-\delta^2\right) \,,
\end{equation}

\begin{equation}
\label{rho2}
\rho\,(\delta) =\frac{\gamma}{8\pi r^2_\Sigma} \left(9-5\delta^2\right)\,,
\end{equation}  

\begin{equation}
\label{Pr2}
P_r\,(\delta)=\frac{3\gamma}{8\pi r^2_\Sigma}\left(1-\delta^2\right) \,,
\end{equation}

\begin{equation}
\label{Pt2}
P_\perp\,(\delta)=\frac{\gamma}{8\pi r^2_\Sigma}\left(\frac{9\gamma\delta^2-3\gamma\delta^4+2\gamma\delta^6-6\delta^2+3}{1-3\gamma\delta^2+\gamma\delta^4}\right)\,,
\end{equation}
with $\delta = \dfrac{r}{r_\Sigma}$ and $\gamma = \dfrac{M_0}{r_\Sigma}<\dfrac{1}{2}$\,.\\

We draw attention to a misconception in the Hernández and Núñez's paper \cite{HN2004}, in which they omitted a constant corresponding to $8\pi\rho_0$, which multiplies the second term of equation (59) of their paper, that we correct here. Moreover, unlike them, we show that in the interval $3/8 \le \gamma \le 7/16$  all hydrostatic equilibrium and energy conditions are satisfied throughout the distribution of matter. In physical coordinates, this gap is equivalent to $5.0625\times10^{26}kg/m \le \gamma \le 5.90625\times10^{26}kg/m$.\\

\begin{table}
\begin{tabular}{ |c|c|c|c|c| }
	\hline
	Neutron Star & Mass ($M_\odot$) & $\gamma$ ($10^{26}kg/m$) & Radius ($km$) & Reference \\ 
	\hline 
	\multirow{2}{*}{J0348+0432} & \multirow{2}{*}{2.01} & 4.00 & 10 & \cite{A2013} \\
	&  & 2.67 & 15 & \\ 
	\hline
	\multirow{2}{*}{J2215+5135} & \multirow{2}{*}{2.27} & 4.52 & 10 & \cite{LSC2018} \\
    &  & 3.01 & 15 & \\ 
	\hline
	\multirow{2}{*}{B1957+20} & \multirow{2}{*}{2.39} & 4.75 & 10 & \cite{M2011} \\
	&  & 3.17 & 15 & \\ 
	\hline
	\multirow{2}{*}{4U 1700-377} & \multirow{2}{*}{2.44} & 4.85 & 10 & \cite{R2011} \\
	&  & 3.24 & 15 & \\ 
	\hline
	\multirow{2}{*}{J1748-2021B} & \multirow{2}{*}{2.74} & 5.45 & 10 & \cite{F2008} \\
    &  & 3.63 & 15 & \\ 
    \hline
    \multirow{2}{*}{Upper TOV limit} & \multirow{2}{*}{3.00} & 5.97 & 10 & \cite{B1996} \\
    &  & 3.98 & 15 & \\ 
    \hline
\end{tabular}
    \label{tab1}
    \caption{Compacteness of some of the most massive neutron stars for typical radius.}
\end{table}

The determination of both mass and radius for neutron stars is not direct. In fact, the measument of the mass depends on the determination of many orbital parameters, not always available. Even for neutron stars with precise masses, the measurement of the radius has been a challenge \cite{L2012}. In table I, we show some of the most massive neutron star ever found, and the theorical upper Tolman–Oppenheimer–Volkoff limit. In order to estimate the compactness $\gamma$ (mass-radius relation) of each star, we collect masses from the literature without taking into account uncertainties, and we adopt the approximate maximum and minimum values for the radius, determined by theoretical models. Observe that the lower limit of the $\gamma$ parameter furnished by our model is the closest to the estimated values. It is important to note that our model refers to a dense object in collapse and not specifically a neutron star.  

\section{Time evolution of the initial solution}

Substituting (\ref{h1}) into (\ref{calor3}), (\ref{Theta3}), (\ref{massa8}), and (\ref{lum9}), respectively, we obtain

\begin{equation}
\label{calor4}
q=\frac{b\gamma\delta}{4\pi r_\Sigma}\left(2\delta^2-3\right) 
\sqrt{1-\gamma\delta^2\left(3-\delta^2\right)}
\left(\frac{f^{1/2}-f^{1/4}}{f^2}\right)  \,,
\end{equation}

\begin{equation}
\label{Theta4}
\Theta=\frac{3b\gamma}{r_\Sigma}
\sqrt{1-\gamma\delta^2\left(3-\delta^2\right)}	
\left(\frac{f^{1/2}-f^{1/4}}{f}\right) \,,
\end{equation}

\begin{equation}
\label{massa10}
m = \gamma r_\Sigma f^{1/2}\left[\frac{2\gamma}{1-2\gamma}\frac{\left( f^{1/2}-f^{1/4}\right)^2 }{f}+1\right] \,,
\end{equation}

\begin{eqnarray}
\label{lum13b}
L = 2\gamma^2 \left(f^{-1/4}-1\right) \left[1+\frac{2\gamma}{1-2\gamma}\left(1-f^{-1/4}\right)\right]^2 \,. 
\end{eqnarray}\

In order to determine if the initial configuration evolves to a black hole, we look for possible solutions to the equation (\ref{red3}) considering the metric (\ref{met12}), that is

\begin{equation}
\label{redB1}
\left[\frac{1}{2}\frac{r\dot{f}_H}{\xi\sqrt{f_H}}+1 \right]_\Sigma = 0 \,.
\end{equation}\

Next, substituting (\ref{fd3}) into (\ref{redB1}), we can obtain the value of $f$, corresponding to the instant of the event horizont formation $t_H$, 

\begin{equation}
\label{redB4} 
f_H = 16\gamma^4 \,.
\end{equation}\

Then, if we choose the minimum and maximum limit for the $\gamma$ parameter, that are $\gamma=3/8$ and $\gamma=7/16$, we have, respectively

\begin{eqnarray}
\slabel{redB5a} 
	f_{H_1} &=& \frac{81}{256} \approx 0.316406 \quad \Rightarrow \quad t_{H_1} \approx -0.236696\,s \,,\\
\slabel{redB5b} 
	f_{H_2} &=& \frac{2401}{4096} \approx 0.586182 \quad \Rightarrow \quad t_{H_2} \approx -0.234751\,s \,.
\end{eqnarray}\

Thus, we can see that the system collapses in a finite time.\\

We can also determine mass-energy when the star crosses the event horizon substituing (\ref{redB4}) in (\ref{massa10}). So,

\begin{equation}
\label{massa11}
m_H=2\gamma^2r_\Sigma \,.
\end{equation}\

Therefore, we can find the mass loss percentage as

\begin{equation}
\label{massa12}
M_L=\left(\frac{M_0-m_H}{M_0}\right) \times 100\%=\left( 1-2\gamma\right) \times 100\% \,.
\end{equation}\ 

The equation (\ref{massa12}) shows that $25\%$ of the initial mass declines for $\gamma=3/8$, and $12.5\%$ for $\gamma=7/16$, which is displayed in figure 1. We could interpret its decrease as loss of matter in the form of radiation. 

Figure 2 shows that an observer at infinity would notice a rapid growth in the brightness of the star, reaching its maximum value, and after decaying until disappearing into the event horizon.

In addition, we can see the heat flux inside the star represented in figure 3, while the figure 4 shows that the expansion scalar decreases during the gravitational collapse, as expected.

The time range analyzed is from the beginning of collapse process ($f\rightarrow 1$) to $f=81/256$ for $\gamma=3/8$, and $f=2401/4096$ for $\gamma=7/16$. Those graphics should be observed from right to the left. 

\begin{figure} 
	\subfloat[\label{fig:B1} $\gamma=\frac{3}{8}$]
	{\includegraphics[width=5.0cm]{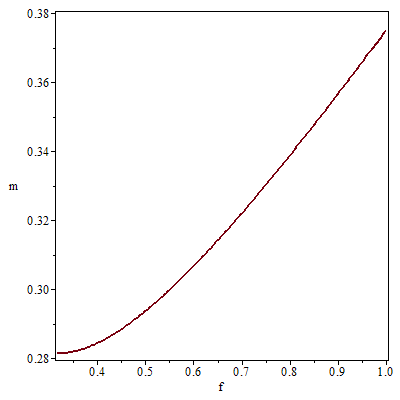}}
	\subfloat[\label{fig:B2} $\gamma=\frac{7}{16}$]
	{\includegraphics[width=5.0cm]{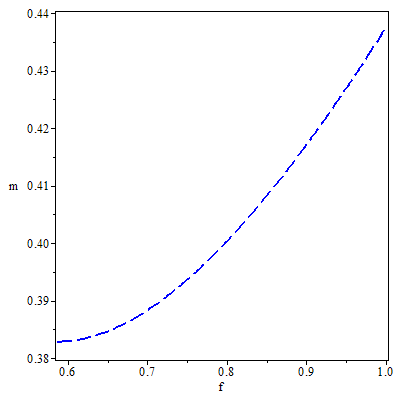}}
	\label{Bmassa}
	\caption{Mass-energy. $m$ is in units of $s$ and $f$ is dimensionless.}
\end{figure}

\begin{figure} 
	\subfloat[\label{fig:B3} $\gamma=\frac{3}{8}$]
	{\includegraphics[width=5.0cm]{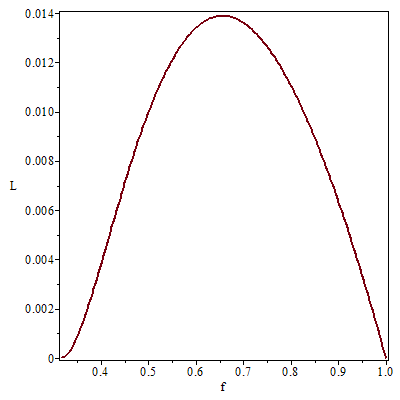}}
	\subfloat[\label{fig:B4} $\gamma=\frac{7}{16}$]
	{\includegraphics[width=5.0cm]{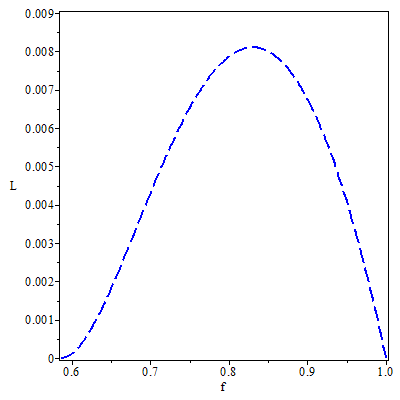}}
	\label{BLuminosidade}
	\caption{Luminosity. $L$ and $f$ are dimensionless.}
\end{figure}

\begin{figure} 
	\subfloat[\label{fig:B5} $\gamma=\frac{3}{8}$]
	{\includegraphics[width=5.0cm]{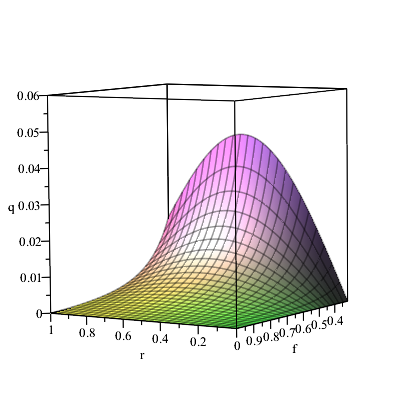}}
	\subfloat[\label{fig:B6} $\gamma=\frac{7}{16}$]
	{\includegraphics[width=5.0cm]{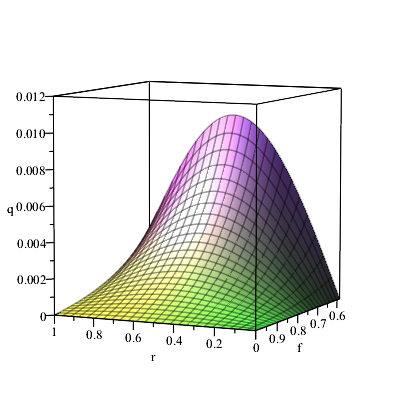}}
	\label{Bcalor}
	\caption{Heat flux. $q$ is in units of $s^{-2}$ and $f$ is dimensionless.}
\end{figure}

\begin{figure} 
	\subfloat[\label{fig:B15} $\gamma=\frac{3}{8}$]
	{\includegraphics[width=5.0cm]{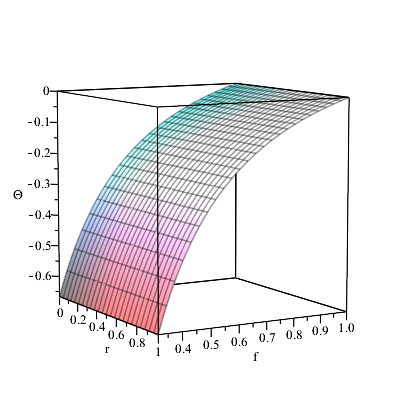}}
	\subfloat[\label{fig:B16} $\gamma=\frac{7}{16}$]
	{\includegraphics[width=5.0cm]{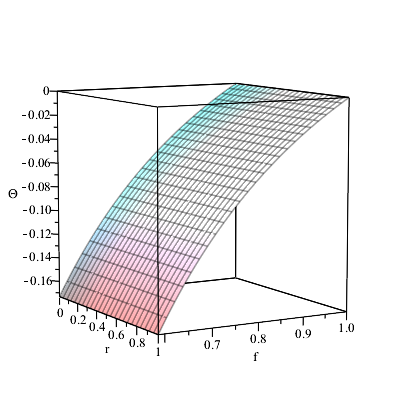}}	
	\label{Bexp}
	\caption{Expansion scalar. $\Theta$ is in units of $s^{-1}$ and $f$ is dimensionless.}
\end{figure}

We can also calculate the effective adiabatic index using (\ref{rho2}) and (\ref{Pr2}), that is, 

\begin{equation}
\label{gamma1}
\Gamma = \left[\frac{\partial(\ln P_r)}{\partial(\ln \rho)} \right]_{r=const} = \left(\frac{\dot{P_r}}{P_r} \right)\left(\frac{\rho}{\dot{\rho}} \right) \,.    
\end{equation}\

Based on the graphics shown in Figs. 5-6, we can say that the bulk viscosity coefficient contributes to the uniformity of $\Gamma$ along the star. This behaviour is more clear in the model with $\gamma=3/8$. 

Although the graphics do not show the behaviour of the effective adiabatic index until the limit $f\rightarrow 1$, we verified that in the presence of bulk viscosity, the effective adiabatic index diverges in all layers of the star in the beginning of the collapse. If we remove the bulk viscosity, this divergence occurs only in the outer layer.

\begin{figure} 
	\includegraphics[width=5.0cm]{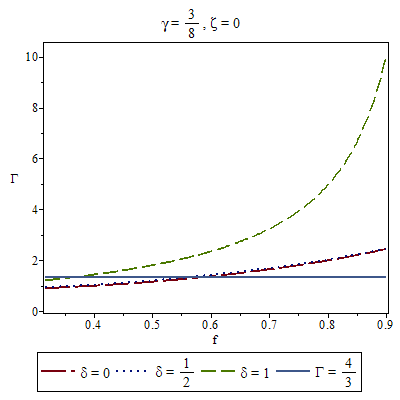}
	\includegraphics[width=5.0cm]{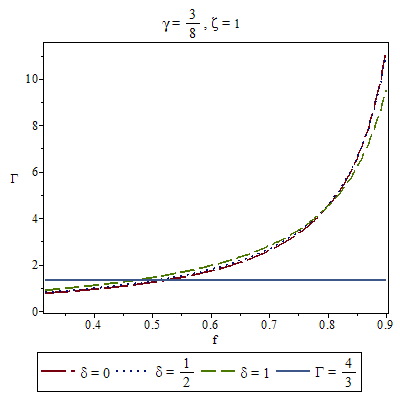}
	\includegraphics[width=5.0cm]{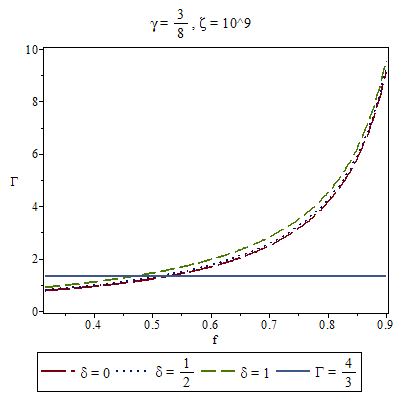}
	\label{Bgamma1}
	\caption{Effective adiabatic index with $\gamma=3/8$. $\Gamma$ and $f$ are dimensionless.}
\end{figure}

\begin{figure} 
	\includegraphics[width=5.0cm]{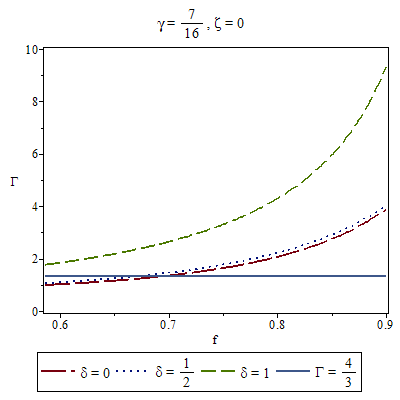}
	\includegraphics[width=5.0cm]{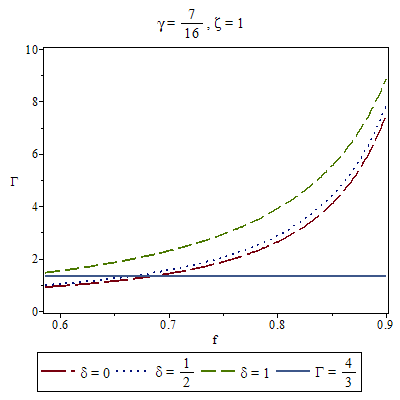}
	\includegraphics[width=5.0cm]{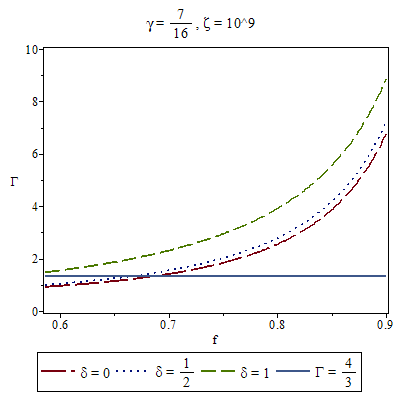}
	\label{Bgamma2}
	\caption{Effective adiabatic index with $\gamma=7/16$. $\Gamma$ and $f$ are dimensionless.}
\end{figure}

\newpage
\section{Energy conditions for an anisotropic fluid with viscosity}

Since the energy-momentum tensor has elements outside of the main diagonal, we must diagonalize it and, to do this, we have to solve the equations  

\begin{equation}
\label{cond1}
\mid T_{\alpha\beta}^{-}-\lambda g_{\alpha\beta}^{-}\mid \,= 0 \,.
\end{equation}\  

The roots of the equation above are the eigenvalues $\lambda$ of diagonalized energy-momentum tensor, and the energy conditions are established in terms of them.

Defining

\begin{eqnarray}
\label{cond2a} 
  P_1 &=& P_r-\zeta\Theta = \frac{1}{8\pi}\left(\frac{h-rh'-1}{r^{2}f}+\frac{h}{4\xi^2}\frac{\dot{f^2}}{f^{2}}-\frac{h}{\xi^2}\frac{\ddot{f}}{f}\right)\,,\\
\label{cond2b} 
  P_2 &=& P_\perp-\zeta\Theta = \frac{1}{8\pi}\left(\frac{h'^{2}-hh''}{2hf}+\frac{h}{4\xi^2} \frac{\dot{f^2}}{f^{2}}-\frac{h}{\xi^2}\frac{\ddot{f}}{f}\right) \,,\\
\label{cond2c} 
  \tilde{q} &=& Bq = \frac{\sqrt{f}}{\sqrt{h}}q\,.
\end{eqnarray}\

Using (\ref{cond1}), we obtain

\begin{equation}
\label{cond5}
(P_2-\lambda)(P_2-\lambda)\left[(\rho+\lambda)(P_1-\lambda)-\tilde{q}^2\right] = 0 \,,
\end{equation}\\*
which results in four solutions

\begin{eqnarray}
\label{cond9} 
\lambda_0 &=&-\frac{1}{2}\left(\rho-P_1+\Delta \right) \,, \\
\nonumber\\
\label{cond10} 
\lambda_1 &=&-\frac{1}{2}\left(\rho-P_1-\Delta \right) \,,
\end{eqnarray}

\begin{equation}
\label{cond12}
\lambda_3 = \lambda_2 = P_2 \,.
\end{equation}\

In the following sections, we introduce the energy conditions for the model proposed here. More details can be found in \cite{KST1988}.

\subsection{Weak energy conditions}

In terms of the eigenvalues, the weak energy conditions can be resumed as

\begin{equation}
\label{fraca1}
-\lambda_{0} \geq 0 \,, 
\end{equation}

\begin{equation}
\label{fraca2}
-\lambda_{0}+\lambda_{i} \geq 0 \qquad (i=1,2,3)\,.
\end{equation}\

Considering (\ref{cond9}), the inequation (\ref{fraca1}) can written as

\begin{equation}
\label{fraca3}
\rho-P_1+\Delta \geq 0 \,.
\end{equation}\

Setting $i=1$ in (\ref{fraca2}), it is necessary to use  (\ref{cond9}) and (\ref{cond10}). So,

\begin{equation}
\label{fraca4}
\Delta \geq 0 \,.
\end{equation}\

Analogously, setting $i=2$, we have to use (\ref{cond9}) and (\ref{cond12}). Thus,

\begin{equation}
\label{fraca5}
\rho-P_1+2P_2+\Delta \geq 0 \,,
\end{equation}\\*
which has an identical result when we set $i=3$.

\subsection{Dominant energy conditions}

The dominant energy conditions imposes

\begin{equation}
\label{dominante1}
-\lambda_{0} \geq 0 \,, 
\end{equation}

\begin{equation}
\label{dominante2}
\lambda_{0} \leq \lambda_{i} \leq -\lambda_{0} \qquad (i=1,2,3)\,. 
\end{equation}\

The inequalities (\ref{dominante1}) and (\ref{fraca1}) are the same, so the inequality (\ref{fraca3}) is included in dominant condition, that is

\begin{equation}
\label{dominante0}
\rho-P_1+\Delta \geq 0 \,.
\end{equation}\ 

Setting $i=1$, and using (\ref{cond9}), (\ref{cond10}), and (\ref{dominante2}), we get

\begin{equation}
\label{dominante4}
0 \leq \Delta \leq \rho-P_1+\Delta \,.
\end{equation}\

The inequality (\ref{dominante4}) implies

\begin{equation}
\label{dominante6}
\Delta \ge 0 \,, 
\end{equation}

\begin{equation}
\label{dominante5} 
\rho-P_1 \ge 0 \,.
\end{equation}

For $i=2,3$, we get

\begin{equation}
\label{dominante10}
\rho-P_1+2P_2+\Delta \ge 0 \,,
\end{equation}

\begin{equation}
\label{dominante11}
\rho-P_1-2P_2+\Delta \ge 0 \,.
\end{equation}

\subsection{Strong energy conditions}

Finally, the strong energy conditions are given by

\begin{equation}
\label{forte1}
-\lambda_{0}+\sum_{i=1}^{i=3}\lambda_{i} \geq 0 \,,
\end{equation}

\begin{equation}
\label{forte2}
-\lambda_{0}+\lambda_{i} \geq 0 \qquad (i=1,2,3)\,.
\end{equation}\

Substituting (\ref{cond9}), (\ref{cond10}), and (\ref{cond12}) into (\ref{forte1}), we obtain

\begin{equation}
\label{forte3}
2P_2+\Delta \ge 0 \,.
\end{equation}\

As (\ref{forte2}) is identical to (\ref{fraca2}), the inequalities are the same, that are

\begin{equation}
\label{forte4}
\Delta \geq 0 \,, 
\end{equation}

\begin{equation}
\label{forte5}
\rho-P_1+2P_2+\Delta \geq 0 \,.
\end{equation}

\subsection{Analysis of the  energy conditions}

The Table 2 summarizes the complet set of energy conditions. There, we can see that the validity of the first inequality and third one implies that second one is satisfied. In the same way, the fifth inequality is fulfilled when the third and sixth ones are. So, the analysis of the first, third, fourth, and sixth ones are enough. Next, we show them in Figs. 7-10 for specific values of $\gamma$, which correspond to the limits that preserve the hydrostatic equilibrium and the energy conditions for initial configuration.\\

\begin{table}[h]\centering
	\begin{tabular}{|c|c|}
		\toprule
		$\Delta \geq 0$ & WEC / DEC / SEC \\ 
		\midrule
		$\rho-P_1+\Delta \geq 0$ & WEC / DEC \\
		\midrule
		$\rho-P_1 \geq 0$ & DEC \\
		\midrule
		$\rho-P_1-2P_2+\Delta \geq 0$ & DEC \\
		\midrule
		$\rho-P_1+2P_2+\Delta \geq 0$ & WEC / DEC / SEC \\
		\midrule
		$2P_2+\Delta \geq 0$ & SEC \\
		\bottomrule
	\end{tabular}
	\caption{Energy conditions}
\end{table}

For the chosen models ($\gamma=3/8$ and $\gamma=7/16$), all the energy conditions are satisfied for any radius and time.

\begin{figure} 
	\subfloat[\label{fig:B7} $\gamma=\frac{3}{8}$]
	{\includegraphics[width=6.0cm]{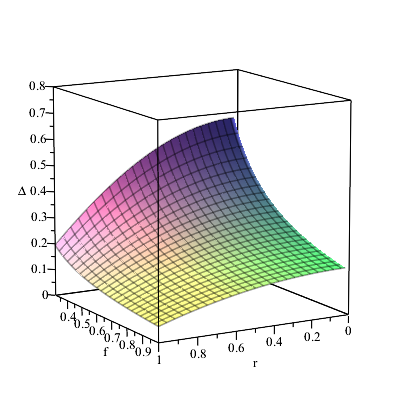}}
	\subfloat[\label{fig:B8} $\gamma=\frac{7}{16}$]
	{\includegraphics[width=6.0cm]{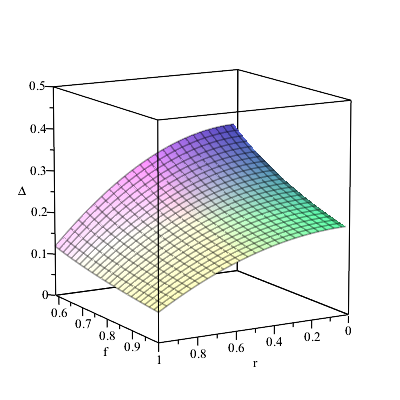}}
	\label{BDelta}
	\caption{$\Delta$. The function $\Delta$ is in units of $s^{-2}$, $r$ is in units of $s$, and $f$ is dimensionless.}
\end{figure}

\begin{figure} 
	\subfloat[\label{fig:B9} $\gamma=\frac{3}{8}$]
	{\includegraphics[width=6.0cm]{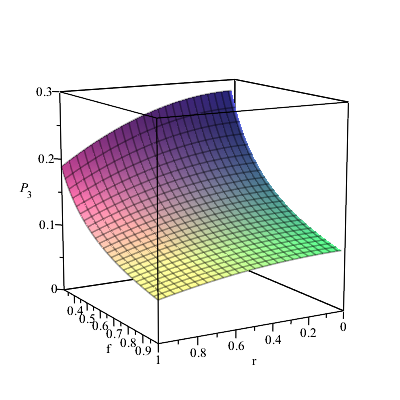}}
	\subfloat[\label{fig:B10} $\gamma=\frac{7}{16}$]
	{\includegraphics[width=6.0cm]{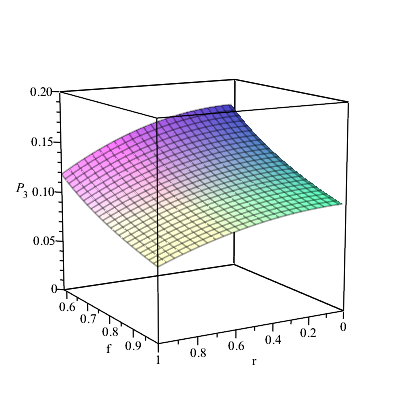}}
	\label{BP3}
	\caption{$P_3=\rho-P_1$. The function $P_3$ is in units of $s^{-2}$, $r$ is in units of $s$, and $f$ is dimensionless.}
\end{figure}

\begin{figure}
	\subfloat[\label{fig:B11} $\gamma=\frac{3}{8}$]
	{\includegraphics[width=6.0cm]{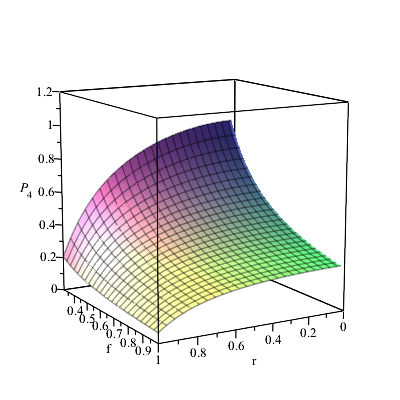}}
	\subfloat[\label{fig:B12} $\gamma=\frac{7}{16}$]
	{\includegraphics[width=6.0cm]{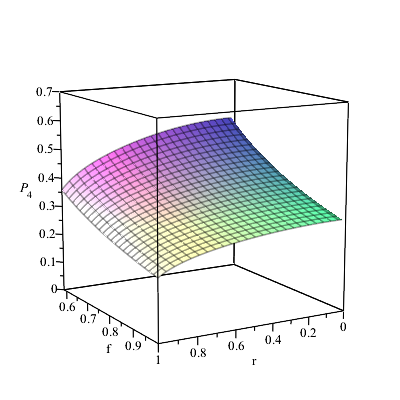}}
	\label{BP4}
	\caption{$P_4=2P_2+\Delta$. The function $P_4$ is in units of $s^{-2}$, $r$ is in units of $s$, and $f$ is dimensionless.}
\end{figure}

\begin{figure} 
	\subfloat[\label{fig:B13} $\gamma=\frac{3}{8}$]
	{\includegraphics[width=6.0cm]{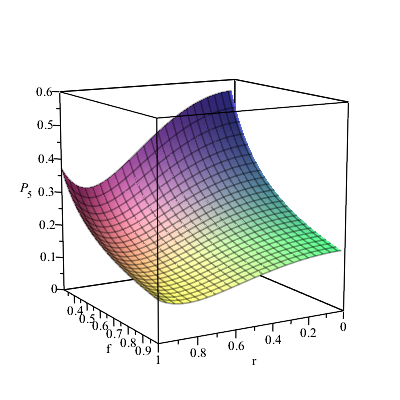}}
	\subfloat[\label{fig:B14} $\gamma=\frac{7}{16}$]
	{\includegraphics[width=6.0cm]{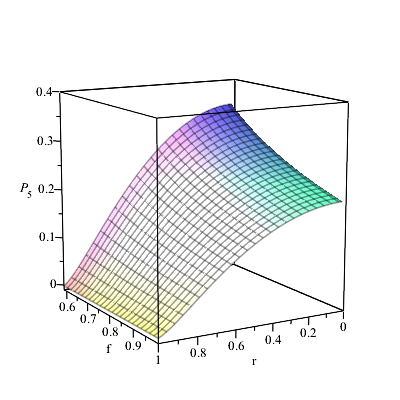}}
	\label{BP5}
	\caption{$P_5=\rho-P_1-2P_2+\Delta$. The function $P_5$ is in units of $s^{-2}$, $r$ is in units of $s$, and $f$ is dimensionless.}
\end{figure}

\newpage
\section{Conclusion}

In this work, we proposed a time-dependent solution for an imperfect fluid with pressure anisotropy which obeys, at least initially, a non-local state equation. 

We admit the possibility of separating variables in order to simplify the field equations and obtain analytic solutions. Moreover, we consider an energy-momentum tensor that includes radial heat flow, anisotropic pressures and bulk viscosity. We restricted our models to $3/8 \le \gamma \le 7/16$ because all the energy conditions are satisfied in this range at the initial time.

The star undergoes a gravitational collapse process, resulting in the formation of an event horizon, when the star becomes a black hole. In this model, although it was not possible to obtain explicity the temporal function present in the metric, and as this function varies between $1$ and $0$ in the gravitation collapse process, it was possible to study some important physical quantities, such as mass-energy, luminosity, heat flux, scalar expansion and the effective adiabatic index. We also found the percentage of the stellar mass lost depending on the initial mass-radius ratio. It is important to point out that all the energy conditions are satisfied during all the stages of the collapse process.

\section*{Acknowledgments}
The financial assistance from Conselho Nacional de Desenvolvimento Cient\'ifico (CNPq), Fundação Carlos Chagas Filho de Amparo à Pesquisa do Estado do Rio de Janeiro (FAPERJ) and Coordenação de Aperfeiçoamento de Pessoal de Nível Superior (CAPES) are gratefully acknowledged. We also would like to thank our colleague Juan M. Z Pretel for helpful discussions and comments about this work.


\end{document}